\documentclass[aps,pra,reprint,groupedaddress]{revtex4-1}

\usepackage{graphicx}
\usepackage{hyperref}

\begin{document}


\title{How to implement decoy-state quantum key distribution for a satellite uplink with 50~dB channel loss}



\author{Evan Meyer-Scott}
\email{emeyersc@iqc.ca}

\author{Zhizhong Yan}

\author{Allison MacDonald}

\altaffiliation{Department of Physics and Astronomy, McMaster University, 1280 Main Street W, Hamilton  ON L8S 4M1, Canada}
\author{Jean-Philippe Bourgoin}

\author{Hannes H\"{u}bel}

\altaffiliation{Department of Physics, Stockholm University, 10691 Stockholm, Sweden}

\author{Thomas Jennewein}
\email{tjennewe@iqc.ca}
\affiliation{Institute for Quantum Computing, University of Waterloo, 200 University Avenue W, Waterloo ON
N2L 3G1, Canada}


\date{\today}

\begin{abstract}
Quantum key distribution (QKD) takes advantage of fundamental properties of quantum physics to allow two distant parties to share a secret key; however, QKD is hampered by a distance limitation of a few hundred kilometers on earth. The most immediate solution for global coverage is to use a satellite, which can receive separate QKD transmissions from two or more ground stations and act as a trusted node to link these ground stations. In this article, we report a system capable of performing QKD in the high loss regime expected in an uplink to a satellite using weak coherent pulses and decoy states. Such a scenario profits from the simplicity of its receiver payload, but has so far considered to be infeasible due to very high transmission losses (40 - 50 dB). The high loss is overcome by implementing an innovative photon source and advanced timing analysis. Our system handles up to 57 dB photon loss in the infinite key limit, confirming the viability of the satellite uplink scenario. We emphasize that while this system was designed with a satellite uplink in mind, it could just as easily overcome high losses on any free space QKD link.
\end{abstract}

\pacs{ 03.67.Dd, 42.50.Ex}

\maketitle

\section{Introduction}
Quantum key distribution (QKD) is the most successful application to arise thus far from quantum information theory \cite{BB84,RevModPhys.74.145}, but it carries the drawback of a distance limitation \cite{springerlink:10.1007/BF00191318,Franson:94,PhysRevLett.96.070502,NJP250k,Liu:10,Takesue2007Quantum-,Ursin2007Entangle,PhysRevLett.98.010504,jennewein2000,tittel2000}: even with future advances, no more than 400~km of direct transmission in optical fibers is expected.
However, quantum repeaters and satellites both have the potential to enable worldwide quantum communication. The former is very appealing with recent promising results \cite{refId}, but
is still in the fundamental research stage.
 Satellite QKD, by contrast, is achievable
by today's satellite and quantum technologies, which already have the required performance \cite{1367-2630-11-4-045017}. In the
most feasible scenario, the satellite acts as a trusted node and performs
consecutive key distributions to two different ground stations allowing a symmetric key sharing between any two locations \cite{alleaume-2007}.
Both a downlink and uplink of photons from/to a satellite have been considered to transmit quantum keys, including much work on daylight and nighttime free space links~\cite{Hughes2000Free-spa,1367-2630-4-1-343}. The downlink is expected to experience lower attenuation, since the uplink beam is much more affected by atmospheric turbulence \cite{1367-2630-11-4-045017}. Nonetheless, an uplink may be more practical since it keeps the complex and power-hungry source of photons on the ground, and permits the use of state-of-the-art sources such as weak coherent pulses, heralded or entangled photons, single photon emitters, and possibly quantum memories. With respect to satellite technology, the uplink is beneficial due to looser telescope-pointing requirements, less demanding opto-mechanics (no precision coupling or fibers), and lower data processing needs. Additionally, all required components for the receiver have flown in space, most notably single-photon detectors \cite{krainak:77801J}. However, the channel loss in an uplink is estimated to be above 40~dB, beyond the capability of current QKD systems, and generally deemed impossible. A free space QKD experiment over 144~km~\cite{Ursin2007Entangle} has been performed in 2007, simulating the conditions for a satellite \emph{downlink}, featuring a large 1~m receiver telescope and a modest 30~dB loss. Experiments have also been performed in optical fiber up to 55~dB channel loss~\cite{Collins2007Low-timi}, employing superconducting single photon detectors whose cryogenic temperatures are impractical for a small space-based receiver. Here we show that a QKD \emph{uplink} in a loss regime beyond 40~dB is indeed feasible by implementing a photonic system which includes an innovative photon source, advanced timing analysis, and commercial single photon detectors with the highest overall figure of merit \cite{Hadfield:2009fk}. Our system can perform QKD based on weak coherent pulses and decoy states up to 57~dB total loss (channel + receiver loss) in the infinite key limit \cite{PhysRevA.72.012326}, and has the potential to overcome finite size effects on a single satellite passage \cite{Sun2009Decoy-st}. Our approach could be implemented immediately in a satellite mission.

In support of our experimental work, we have performed a rigorous analysis of channel performance for satellite uplinks and downlinks, including diffraction, atmospheric turbulence in the Hufnagel-Valley model~\cite{Smith1993The-Infr}, pointing error, atmospheric absorption~\cite{anderson:2}, multiphoton statistics, telescopic losses, detector efficiencies, satellite orbit statistics, and background noise, to produce secure key rate statistics for a variety of conditions and systems. As a specific example, for an uplink to a satellite 600~km high, using a 25~cm diameter telescope on the ground and 30~cm on the satellite, our model shows about 80\% of total satellite passages over the ground station will be usable for QKD asymptotically (infinite key limit), with an average total loss of 52~dB, an order of magnitude beyond the capability of current QKD systems.

\section{Technology considerations for satellite transmission}
The most obvious challenge in a satellite uplink is the sheer link distance: it can be 500~km to more than 30,000~km depending on the satellite orbit, making the quantum channel extremely lossy. Additionally, noise due to detector dark counts and stray light, especially moonlight and terrestrial light, will make satellite QKD more demanding. Finally, the short duration of each satellite passage, on the order of hundreds of seconds, makes proving security of QKD difficult, given the small number of quantum signals received. To address the challenges of a satellite uplink, both physical and technical parameters must be tuned. The first variable that can be chosen to minimize loss is the wavelength of the photons. Beam spread due to diffraction is one of the main sources of loss and is proportional to wavelength, so short wavelength photons are preferred. After considering the optical transmission of the atmosphere and single photon detector capabilities, a good choice is $\lambda= 532$~nm, which
enables the use of thin silicon avalanche photodiodes \cite{Rech2004Photon-t}. This type of detector has the
highest figure of merit for single photon quantum information
applications \cite{Hadfield:2009fk}, based on efficiency, timing jitter and dark count rate. In order to limit background noise, the system must employ short pulses and temporally precise detection which allow temporal filtering of received signals.  The optimization of this temporal filtering is described below. Furthermore, a high system clock rate is important to generate enough signals to account for statistical fluctuations in estimation of an eavesdropper's information (finite size effects). As a final consideration, the QKD system must have phase
randomization such that subsequent pulses share no phase relation, which is assumed in security proofs to limit information given to an eavesdropper. 
\section{System configuration}

\begin{figure*}[htp]
\includegraphics[width=2\columnwidth]{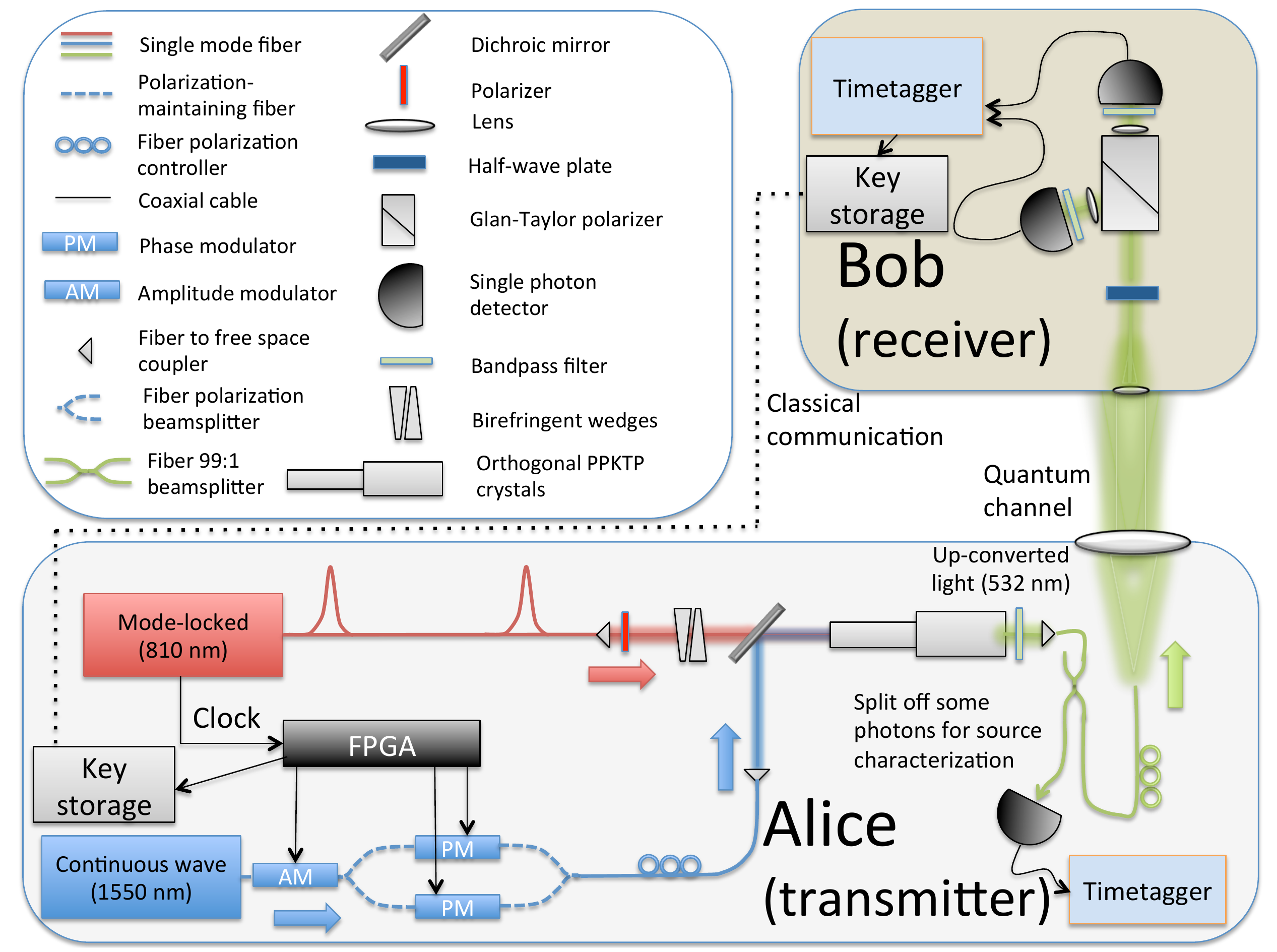}
\caption{Simplified schematic of QKD system for high loss link. Alice's up-conversion photon source produces photons at 532~nm which are sent through the controllable-loss channel to Bob's receiver.}
\label{fig_system}
\end{figure*}

Our weak coherent pulse decoy-state system satisfies all the above requirements through the
sum-frequency generation, or up-conversion method of photon production in a $\chi^{(2)}$ nonlinear crystal. The design and
implementation are illustrated in Fig. \ref{fig_system}.
To provide short pulses and fast modulation, light from a mode-locked titanium
sapphire laser at 810~nm is combined in two type-I Periodically-Poled KTP crystals
with light from a 1550~nm continuous-wave laser to produce, due to energy conservation, photons at 532~nm. The arrangement is equivalent to an asymmetric down-conversion entanglement source run in reverse \cite{Hubel:07}, and uses two orthogonally-oriented PPKTP crystals, for polarization-insensitive up-conversion. The source employs phase precompensation using birefringent wedges in the 810~nm beam to compensate for temporal walkoff in the PPKTP crystals. The pulsed 810~nm beam is set to 45$^\circ$ polarization (coherent superposition of horizontal and vertical), while the 1550~nm pump light is modulated in polarization (qubit state) and intensity (signal or decoy). In this configuration, the output pulses at 532~nm follow the pulse length of the 810~nm  laser and the polarization and amplitude of the modulated 1550~nm beam. The modulation is accomplished with off-the-shelf telecom waveguide modulators, which show high stability and switching contrast, and switching speeds of a few GHz. The power of the two input beams is controlled such that the output pulses at 532~nm contain around one photon per pulse, as determined by the optimal average photon number for decoy-state QKD. The phase randomization
is also accomplished with the telecom laser, whose coherence time ($\sim$5~ ns) is
less than the period between adjacent pulses emitted from the
mode-locked laser.

The up-converted photons are collected into single mode fiber, then Alice splits off some photons with a 99:1 fiber beamsplitter for source characterization. The light is allowed to exit at the fiber tip to the quantum channel, then passes through an adjustable lens to control the beam size at Bob's receiver and therefore the channel loss. Bob's lens selects a small portion of the beam to simulate a high loss channel to space. Bob performs active basis choice with a half-wave plate~\footnote{The optimal solution for this short wavelength is to use passive basis choice, wherein the incoming photon passes through a 50:50 beamsplitter and is measured in either the rectilinear or diagonal basis depending on the path randomly taken out of the beamsplitter. Because only two detectors were available for this experiment, we implemented a slow, active basis choice: a motorized rotation stage containing a half-wave plate was rotated to $0^\circ$ or $22.5^\circ$ to choose to measure in the rectilinear and diagonal bases respectively.}, then the light passes through a polarizer (to determine the bit value) and narrow-band filters before arriving at silicon single-photon detectors from Micro Photon Devices. The detectors have a peak efficiency of 48\% at 550 nm, 10 dark counts per
second and 30~ps timing resolution, which allows temporal exclusion of much background noise. Given a minimum total loss of 40~dB and an achievable 1~GHz clock rate, the maximum count rate seen by each detector is around 20,000 counts per second, making losses due to dead time (70~ns) negligible. The detector events are registered and digitized using a timetagging module with 156~ps resolution. All these components are
commercially available and many are already space qualified or undergoing qualification, making this system practical for satellite
applications.

\section{Decoy-state protocol}

Weak coherent pulse (WCP) sources based on (up-conversion of) highly attenuated lasers are attractive for QKD; however, because of the Poissonian statistics of photon number in laser pulses, some pulses will have more than one photon and be subject to the photon number splitting attack \cite{1367-2630-4-1-344}. In this attack, an adversary Eve splits off one photon from the pulse and stores it to measure only after the legitimate party Bob reveals his measurement basis. Eve then measures in the correct basis, and so gains full information about multi-photon pulses without leaving a trace. To combat this attack, the decoy-state protocol was introduced~\cite{PhysRevLett.91.057901,PhysRevLett.94.230504}, wherein Alice changes the average photon number of randomly interspersed pulses from the signal level $\mu$ to the decoy level $\nu$. Since Eve cannot know whether a given pulse is a signal or decoy pulse, the decoy pulses allow much better bounds on how much information Eve has gained from multiphoton signals, and thus how much privacy amplification must be performed. The asymptotic key rate (adapted from Ref. \cite{PhysRevA.72.012326}) per laser pulse obtainable from a decoy pulse protocol is

\begin{equation}
R\ge q\frac{N_\mu}{N_\mu+N_\nu} \{-Q_\mu f(E_\mu) H_2(E_\mu) + Q_1 \left[1-H_2(e_1)\right]\},
\label{eqn_asymp}
\end{equation}

 where $q=1/2$ is the basis reconciliation factor, $Q_\mu$ is the signal gain, i.e. the ratio of Bob's detections to pulses sent by Alice for average photon number $\mu$, $E_\mu$ is the quantum bit error rate for signal pulses, $f(E_\mu)=1.22$ is the error correction efficiency for practical error correction codes, $H_2(x)=-x\log_2(x)-(1-x)\log_2(1-x)$ is the binary entropy function, and $Q_1$ and $e_1$ are the estimated gain and error rate for single photon pulses. The factor $\frac{N_\mu}{N_\mu+N_\nu}$ is added since only detections of the signal state $\mu$ contribute to the final key, and $N_{\mu/\nu}$ is the number of signal/decoy detections. The key rate is then the gain of single-photon pulses, less the error correction on all signal pulses, less the privacy amplification on single-photon pulses. Note that this key rate should be multiplied by the laser pulse rate to obtain secure key bits per second.

 We chose the two-decoy protocol from Ref. \cite{PhysRevA.72.012326}. In this protocol, Alice sends randomly a signal pulse with average photon number $\mu$, a decoy pulse with average photon number $\nu<\mu$, or the vacuum. In our case, to illustrate the utility of the two-decoy method, we took the vacuum as being sent between adjacent laser pulses. $Q_1$ and $e_1$ were therefore calculated from Section D of Ref. \cite{PhysRevA.72.012326}, allowing a final secure key rate from equation (\ref{eqn_asymp}).

\section{Timing synchronization}
\begin{figure}[htp]
\includegraphics[width= \columnwidth]{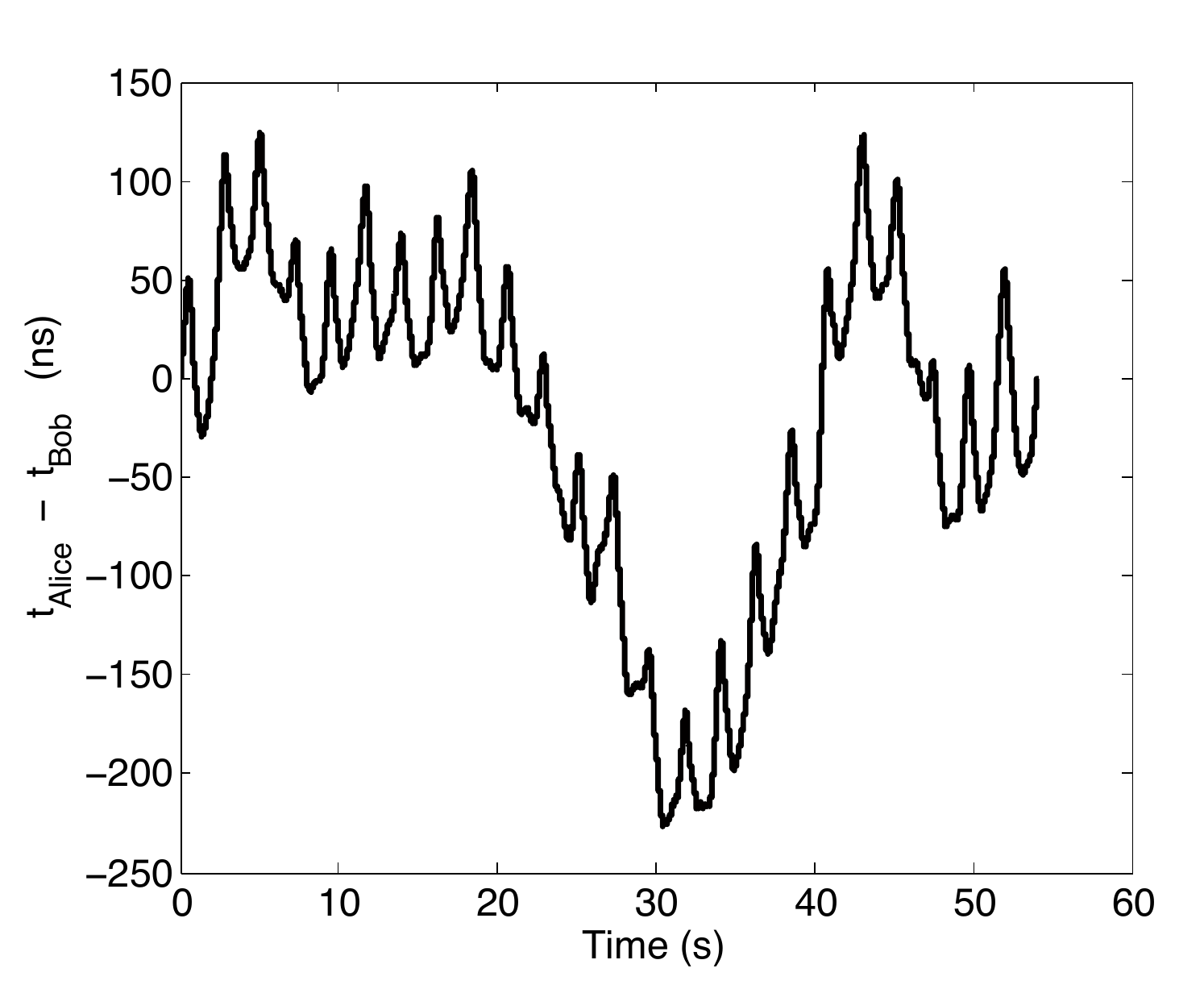}
\caption{Typically observed drift between Alice's and Bob's clocks. Alice's clock is determined by the repetition rate of the mode-locked laser and Bob's comes from his timetagger. Drifts in the clock are large compared to the nominal laser clock period of 13~ns, making timing synchronization a necessity. Our synchronization scheme correctly aligns Bob's detection events independent of which device is drifting.}
\label{fig_drift}
\end{figure}

Since the clock periods at Alice and Bob will inevitably drift, a synchronization based on the sent and received signals had to be devised. As seen in Fig. \ref{fig_drift}, the timetagging clock at Bob may drift hundreds of nanoseconds relative to the laser clock (period $=13$~ns) at Alice. As an example, if the laser clock period is shortened by only 1~fs, the clocks will be offset by 76~ns after one second, making signal identification impossible. Therefore, timing synchronization between Alice and Bob is necessary, and is accomplished here by timetagging a frequency-divided version of the laser clock, following rough ($\sim$ns) alignment with GPS signals. Additionally, a known pseudorandom sequence could be inserted to allow absolute time alignment if required~\cite{PhysRevLett.98.010504}. Bob then sends his timetags to Alice who uses her timetagged laser clock signal to stretch or compress portions of Bob's detection timetags depending on the fluctuations as caused by cavity length changes in the laser or drifts in the timetagger's clock. Thus Alice can identify which tags to keep based on timing  and relay this information to Bob. This could be performed over the classical communication channel, and since only detection times and not bit or basis values are revealed, no information is leaked to Eve. In the satellite application, we must also consider the fast movement of the satellite towards or away from the base station. It becomes necessary to know precisely the position of the satellite through orbit analysis and direct time-of-flight measurements, which provide a smoothly varying ground-satellite time-of-flight function~\cite{1367-2630-10-3-033038}. In post processing, Alice can then align Bob's timetags every second by applying the smooth time offset and searching for a coincidence peak with her laser clock's timetags. We estimate that with a 1~GHz clock rate and the worst case of 57~dB total loss, Bob would receive approximately 800 legitimate signals each second, more than sufficient for the time synchronization procedure.

\section{Quantum Key Distribution performance}
\begin{figure*}[htp]
\includegraphics[width= 2\columnwidth]{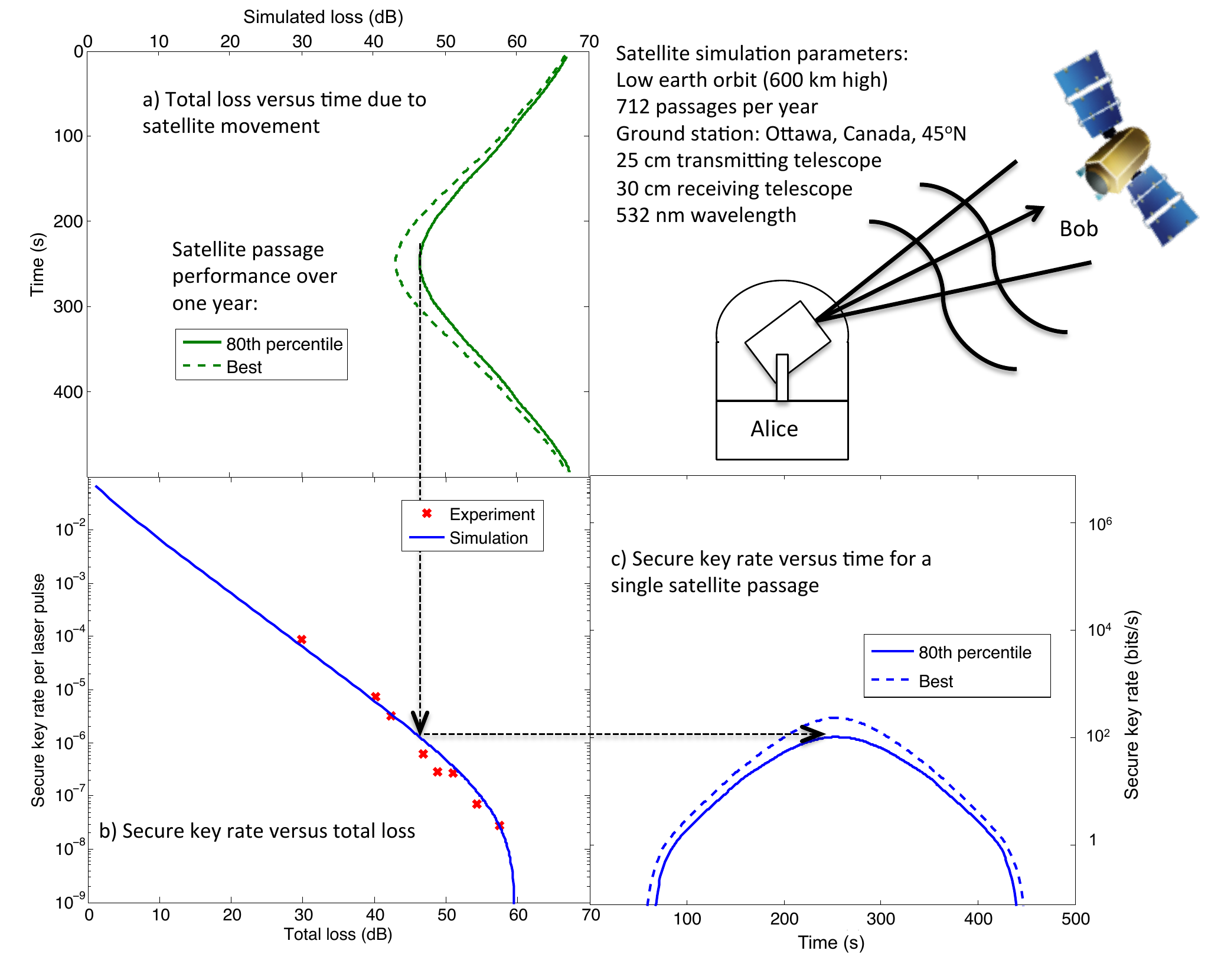}
\caption{Secure key rate over high loss channel. a) Simulation of total loss versus visible passage time for a satellite uplink. Here we show simulation data for the best overall passage in that year and, for comparison, the 142nd best passage, i.e. 80th percentile of the 712 total passages for the year. The loss is minimum as the satellite is closest to the ground station (highest elevation angle) and increases as the satellite approaches the horizon. b) Experimental results and simulation of secure key rate versus loss. Our data agree well with the
theoretical curve, which uses a quantum optical simulation to predict key rates. Deviations from the theoretic curve are caused by imperfect alignment of the polarization inside the optical fibers. Treatment of error analysis is included in QKD security proofs, and is generally based on upper bounding the information given to an eavesdropper compatible with measurement results. c) Expected secure key rate versus time for a satellite passage, based on simulations and experimental parameters. The secure key rate in bits/s on the right axis assumes the 76~MHz clock rate of our source. The loss versus time and secure key rate versus loss curves combine to produce the output key rate over one satellite passage. }
\label{fig_keyrate}
\end{figure*}

Our main results are summarized in Fig. \ref{fig_keyrate}. Using our experimental setup, the detection rate and quantum bit error rate (QBER $= E_\mu$) in each of the rectilinear and diagonal bases for signals and decoys were measured and a final secure key rate from equation (\ref{eqn_asymp}) was calculated. A pseudorandom sequence of 256 pulses was repeated and the resulting timetags formed into a histogram to give information on each individual pulse state, allowing full characterization of the system's capability. The results versus loss in Fig. \ref{fig_keyrate}b, based on many 1000~second data collection runs at a clock rate of 76~MHz, show secure key distribution is possible up to 57~dB experimentally. The secure key generation rate at this maximum 57~dB is 2~bits/s, highlighting the viability of the quantum optics and detectors required for this high loss. Allowing 6~dB for receiver and detector efficiency, this permits channel losses up to 51~dB, higher than any WCP systems previously built \cite{Takesue2007Quantum-,NJP250k,Liu:10}. 

We additionally performed a quantum optical simulation including photon production, channel transmittance and detection, to predict a secure key rate versus total loss as shown in Fig. \ref{fig_keyrate}b. The simulations show key generation is possible up to  59~dB and agree with the experimental results.

To highlight the viability of our system for a satellite uplink, simulations of satellite orbits over one year were performed to predict the total channel loss versus passage time of the satellite. Using a realistic orbit at 600~km height, 712 satellite passages over our hypothetical ground station near Ottawa, Canada were predicted, about 80\% of which have a portion with low enough loss for QKD (excluding cloudy nights). The total loss versus time of the overall best single passage and of the 80th percentile passage are plotted in Fig. \ref{fig_keyrate}a. The average total loss for usable passages for QKD is 52~dB, well within the capability of our system.

\begin{figure}[htp]
\includegraphics[width=\columnwidth]{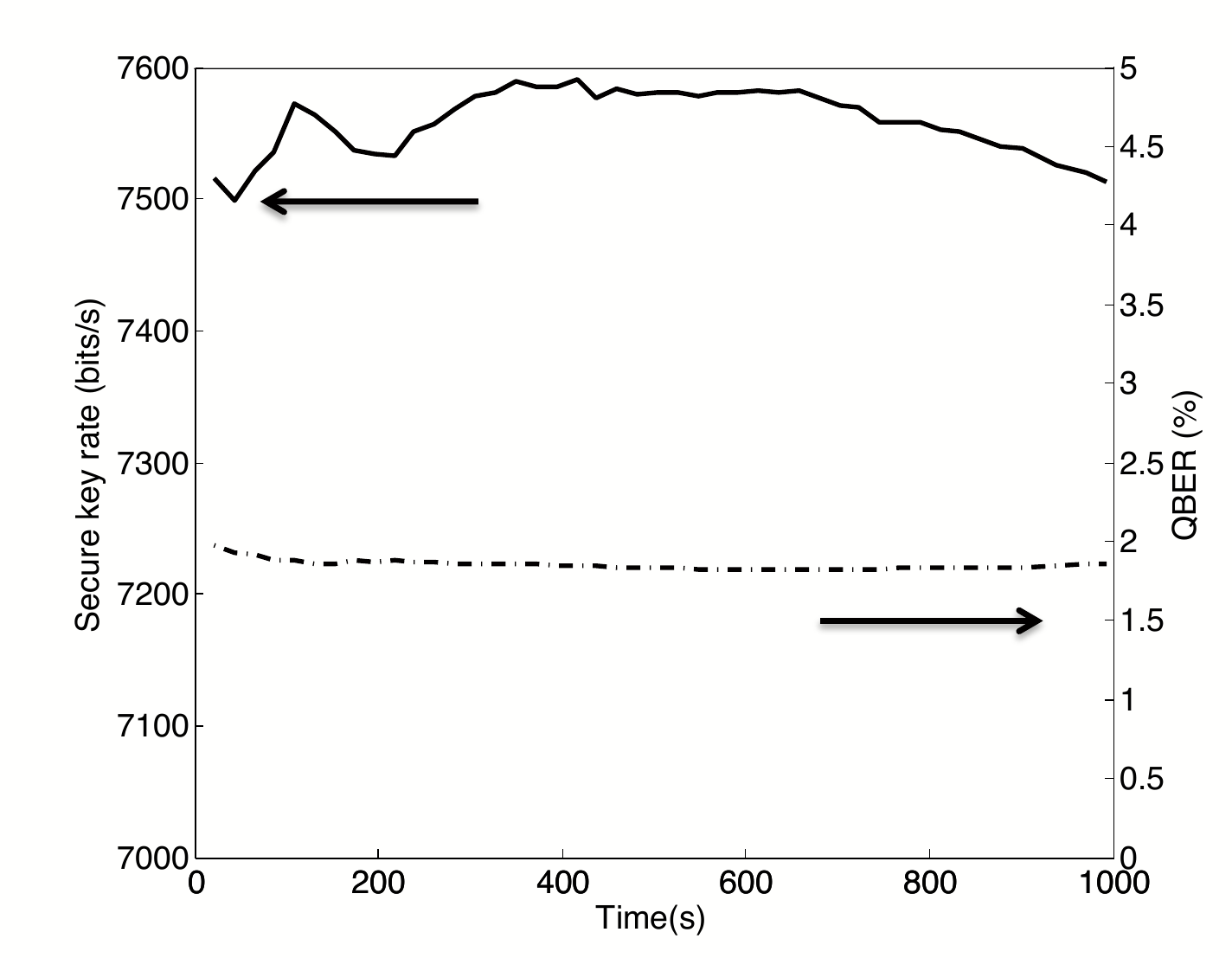}
\caption{Stability of secure key rate (solid line) and QBER (dashed line). Data were taken at 30~dB total loss over 1000~s, and the mean secure key rate and QBER are $7560\pm24$ bits/s and $1.85\pm 0.03~$\% respectively. The small drifts are likely due to temperature fluctuations which alter the polarization transformation in the connecting optical fibers.}
\label{fig_skovertime}
\end{figure}

Finally, the total loss versus time for a satellite passage and secure key rate versus total loss can combine to produce Fig. \ref{fig_keyrate}c, secure key rate versus time for a satellite passage. The rate is given in bits per laser pulse on the left axis and bits per second on the right, based on our clock rate of 76~MHz. The curves in Fig. \ref{fig_keyrate}c can be integrated to find total bits of secure key generated over one passage. For the 80th percentile passage shown here, a total of $5.7\times10^4$ bits of secure key could be generated with our 76~MHz system. Additionally, as shown in Fig. \ref{fig_skovertime}, our photon source is sufficiently stable for key generation without active feedback for the duration of a satellite passage.

When statistical fluctuations \cite{Sun2009Decoy-st} and information theoretic security proofs \cite{PhysRevLett.100.200501} are considered, our simulations predict about 55~dB average loss is permissible over a single satellite passage with our hardware and an achievable 1~GHz clock rate. Recent work on the finite-key problem for qubits \cite{2011arXiv1103.4130T} should allow channel losses to be extended further (once optical modes are considered), making a higher number of yearly satellite passages usable for QKD. 

\section{Removal of background noise}

\begin{figure*}[htp]
\includegraphics[width= 1.8\columnwidth]{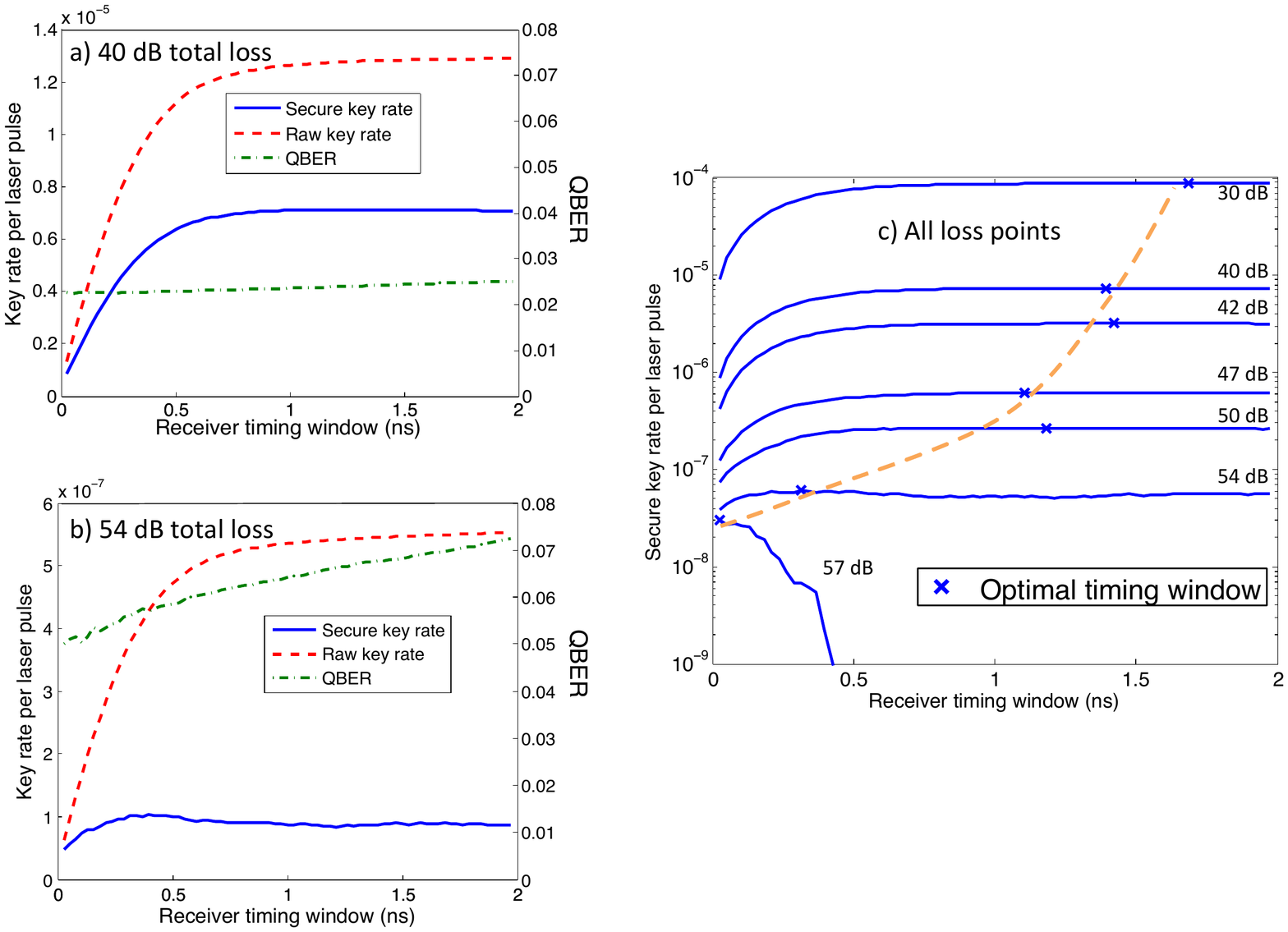}
\caption{Raw key rate (all signal detection events), secure key rate, and quantum bit error rate (QBER) versus timing window from experimental data. a) At 40 dB total loss, b) At 54 dB total loss. Note that raw key rate and QBER both increase with timing window, as more dark and background counts are admitted. Secure key rate shows a maximum at 1.2 ns for the 40~dB case and 0.4~ns for the 54~dB case, as the benefit of increasing the raw key rate is offset by the detriment of increasing the QBER. c) Secure key rate versus timing window for various loss points. The observed optimal timing window is marked on each curve with an X, and the dashed line is a guide to the eye for the optimal window width trend.}
\label{fig_window}
\end{figure*}

To separate legitimate detections from background noise, all detections were timetagged (see Appendix for discussion of timing system), subdivided into 10~ms long sections and then binned with a bin width equal to a fraction of the laser clock cycle. Then the detections from the QKD source should be tightly peaked around the laser pulse times with a width determined by the jitter~\footnote{The final jitter of Alice's mode-locked laser timetags is about 300~ps. Given the 156~ps resolution of the timetagger, and jitter of Alice's electronics of 200~ps, this leaves 160~ps for jitter of the laser. Therefore, improvement is possible most easily in the laser stability and timetag resolution, making this timing method even more appealing.}, and the background noise distributed randomly. The true signals were separated from dark counts and stray light by choosing an optimal window width around the peaks, which narrows with increasing loss as more background counts must be excluded to maintain an acceptable QBER (Fig. \ref{fig_window}). Only timetags within the window contribute to the final key calculations, and those outside are discarded. The number of erroneous background counts per second that are included in the raw key is given by

\begin{equation}
C_{errors}=W\times r\times C_{bkgd},
\end{equation}

where $W$ is the timing window, $r$ is the laser repetition rate, and $C_{bkgd}$ is the total number of background counts per second.  The final key rate given by equation \ref{eqn_asymp} depends on both QBER and raw key rate, leading to the optimal timing windows as noted in Fig. \ref{fig_window}c. The optimal timing window decreases from 2~ns at low loss, to 40~ps at 57~dB, making full use of the good timing afforded by our system, and displaying the linear drop in QBER due to $C_{errors}$ as $W$ narrows.

\section{Discussion and conclusion}

We have demonstrated the design and viability of
a QKD system capable of operation under ultra-high channel losses of up to 57~dB. Our system therefore satisfies the challenging requirements for uplink of quantum keys to a satellite, and future improvements will allow the technical requirements to be satisfied with full information theoretic security. As noted above, a space-based quantum receiver is less demanding than a quantum source, as all required components for the receiver have flown in space \cite{Prochazka2009Photon-c}, so a near-term satellite mission using our approach as a prototype is possible.

 The still outstanding challenges for a full-scale QKD satellite mission include determining sufficient classical processing and communication
bandwidth and designing a reference frame system to compensate for
the slow rotation of the satellite
\cite{1367-2630-11-6-065004,PhysRevA.82.012304}, which will be our next steps
of research. Additionally, how best to deal with a strongly fluctuating channel while maintaining security is an open question \cite{1367-2630-11-4-045017,PhysRevA.84.032340,2011arXiv1110.1440V}. As a first step, we simulated secure key rate versus loss, comparing a static channel to a fluctuating channel with log-normal probability density function~\cite{1464-4266-6-8-018}. We found no difference in secure key rate even for strongly turbulent atmosphere, so long as the \emph{average} channel loss was the same. 
Additionally, putting tighter bounds on the finite-key problem for realistic implementations is a great challenge for theorists, to enable the use of lossier channels than ever \cite{2011arXiv1103.4130T,PhysRevA.83.022330}. Our system must be updated to include a truly random pulse sequence on Alice's side, and a receiver on Bob's side capable of measuring in both rectilinear and diagonal bases simultaneously with passive basis choice. To bring our system to the desired 1~GHz clock rate is not difficult, as the modulators can handle a few GHz and the upconversion process is clock rate independent. We would simply require a mode-locked laser with shorter cavity length and updated electronics.

Finally, with a quantum receiver in space and a suitable photon source, a number of additional quantum physics experiments over
ground-space distance become viable, including teleportation and
entanglement swapping \cite{Aspelmeyer2003Long-dis}, fundamental tests of quantum mechanics \cite{ANDP:ANDP831}, and tests of new physical theories
\cite{RMD09}. In addition, an entangled photon source which emits one photon around the desired 532~nm is envisaged for the future \cite{Spillane2007Spontane,PhysRevA.81.031801}. This photon would be directed to the satellite while the other photon of the entangled pair would be in the telecom band around 1550~nm, suitable for long-distance transmission in optical fibers. A central ground station containing the source could be connected locally by fibers to end-users, and globally via satellite to another such ground station.  Furthermore, it is possible that the uplink transmission can be enhanced by implementing wave-front corrections of the transmitted optical beam, through adaptive optics \cite{berkefeld:77364C}. This technology is used in astronomic observation, and could be realized at the ground station even once the mission is deployed. In summary, the future for QKD using satellites is bright, the uplink is demonstrably feasible, and in the near term we expect to see multiple satellite missions for quantum information, both for fundamental science and applications.

\appendix*
\section{Receiver timing system}
In practical QKD systems, the use of timing information is necessary to exclude illegitimate detections \cite{Collins2007Low-timi}. Our system employs free-running detectors and timetags every detection event, in contrast to gated detection schemes which only open detectors during the specified arrival time of a pulse. Both are subject to detector control attacks \cite{1367-2630-11-6-065003,1367-2630-13-1-013043,2011arXiv1101.5289W} with most effort being focused on gated avalanche photodiodes for telecom applications \cite{Lydersen2010Hacking-}. For satellite applications, the timetagging method is preferable as it requires much less data transmission. A recent precise timing experiment \cite{Restelli2010Improved} required every gate pulse to be sent classically in parallel to the quantum channel from Alice to Bob, a vast overhead which is impractical for space applications, due to both fast and slow changes in optical path length. By contrast, using the timetagging approach, Bob can send back to Alice only the timetags generated by his receiver, which will be small in number due to the high channel loss. Alice can then align them to her source rate and tell Bob which to keep. As an example,  with a clock rate of 1~GHz Alice would have to send $10^9$ gate pulses per second independent of the loss for gated operation, while if using timetags, Bob would have to transmit only about 5000 timetags per second back to Alice for 50~dB total loss.
\begin{acknowledgments}
The authors thank NSERC (CGS, QuantumWorks, Discovery, USRA), Ontario Graduate Scholarships, CIFAR, Canadian Space Agency, Ontario Ministry of Research and Innovation, and CFI for funding.
\end{acknowledgments}


%
%

\end{document}